\documentclass[prl,twocolumn,showpacs,preprintnumbers]{revtex4-1}

\usepackage{graphicx}
\usepackage{epstopdf}
\usepackage{array,multirow,amsmath,amsfonts}
\usepackage{color}

\newcolumntype{C}[1]{>{\centering\let\newline\\\arraybackslash\hspace{0pt}}m{#1}}

\begin{document}
\preprint{}

\title{
Substrate-tuning of correlated spin-orbit oxides
}

\author{Bongjae Kim$^1$}
\thanks{Present address:
University of Vienna, Faculty of Physics and Center
for Computational Materials Science, Sensengasse 8, A-1090 Vienna, Austria}
\author{Beom Hyun Kim$^1$}
\thanks{Present address: RIKEN, Wako, Saitama 351-0198, Japan}
\author{Kyoo Kim$^{1,2}$}
\author{B. I. Min$^1$}
\email[]{bimin@postech.ac.kr}
\affiliation{$^1$ Department of Physics, PCTP,
Pohang University of Science and Technology, Pohang 790-784, Korea\\
$^2$c\_CCMR, Pohang University of Science and Technology, Pohang 790-784, Korea}
\date{\today}

\begin{abstract}
 We have systematically investigated substrate-strain effects on the electronic structures
of two representative Sr-iridates, a correlated-insulator Sr$_2$IrO$_4$ and a metal SrIrO$_3$.
Optical conductivities obtained by the \emph{ab initio}
electronic structure calculations reveal that the tensile strain shifts
the optical peak positions to higher energy side with altered intensities,
suggesting the enhancement
of the electronic correlation and spin-orbit coupling (SOC) strength in Sr-iridates.
The response of the electronic structure upon tensile strain is
found to be highly correlated with the direction of magnetic moment,
the octahedral connectivity, and the SOC strength,
which cooperatively determine the robustness of $J_{eff}$=1/2 ground states.
Optical responses are analyzed
also with microscopic model calculation and compared
with corresponding experiments.
In the case of SrIrO$_3$, the evolution of the electronic structure
near the Fermi level shows high tunability of hole bands,
as suggested by previous experiments.
\end{abstract}


\maketitle

\section{Introduction}

 With recent developments of epitaxial growth technique, substrate-strain engineering
has been employed as a very efficient route to control various physical parameters,
especially in ABO$_3$ transition metal perovskite systems.
As the substrate strain modifies the connectivities of the BO$_6$ octahedra,
such as bond length and bond angle, there occur corresponding macroscopic changes
in the symmetry, electronic structure, and magnetic properties \cite{Rondinelli11,Schlom07}.

 Recently, 5$d$ oxide systems have emerged
as interesting target systems for the substrate engineering.
Due to the prominent role of large spin-orbit coupling (SOC) of
Ir 5$d$ electrons,
which is comparable to the strengths of Coulomb correlation ($U$) and bandwidth ($W$),
intensive attention has been focused on the Sr-iridates
of Ruddlesden-Popper type, Sr$_{n+1}$Ir$_{n}$O$_{3n+1}$.
In particular,
Sr$_2$IrO$_4$ (214) with $n=1$  and SrIrO$_3$ (113) with $n=\infty$
are representative systems, which correspond
to the insulating and metallic limits, respectively.
Both systems have been described based on the $J_{eff}$=1/2 ground states,
where the former has a well-separated Mott-gap,
while the latter is thought to be a correlated metal \cite{BJKim08,Moon08,BJKim09}.
The tetragonality of the system, which is a typical tunable parameter
in the substrate-strain engineering, is found
to be closely correlated to the magnetic-moment direction in 214 systems
which have robust $J_{eff}$=1/2 electronic structure \cite{Jackeli09}.
However, for large tetragonal splitting,
the deviation from the $J_{eff}$=1/2 ground state has been reported
in the recent experiment,
casting the question on the range and the condition
of the $J_{eff}$=1/2 picture \cite{Sala14}.

\begin{figure}[b]
\begin{center}
\includegraphics[angle=0,width=0.4\textwidth]{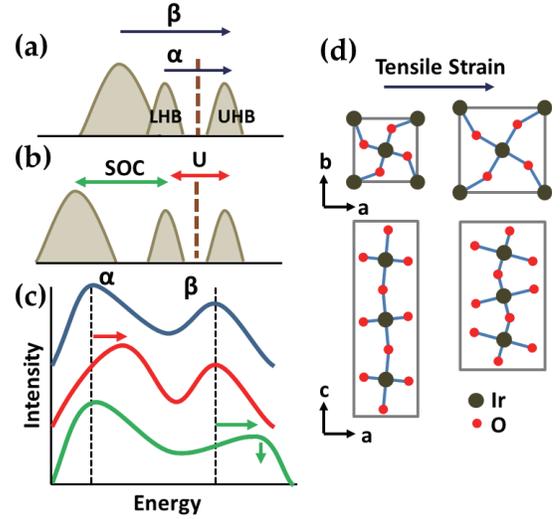}
\caption{
\textbf{Schematic electronic structures and octahedral connectivity of Sr-iridates.}
(a) Schematic band diagrams of Sr$_2$IrO$_4$.
Optical peaks ($\alpha$ and $\beta$) appear due to the transitions represented by arrows.
(b) $U$ determines the gap between LHB and UHB, while SOC separates $J_{eff}$=1/2 and $J_{eff}$=3/2 state.
(c) The roles of $U$ and SOC are schematically drawn. Increasing $U$ and SOC shift
$\alpha$ and $\beta$ peaks, respectively, to high energy sides.
(d) Tensile strain effects on the IrO$_6$ octahedral connectivity of the systems.
For 214 systems, due to its layered structure, there is no octahedral connection apically.
}
\label{fig1}
\end{center}
\end{figure}

Optical experiments for aforementioned Sr-iridates,
which act as a direct probe of electronic structures,
show two-peak structure near the Mott gap region
(see Fig.~\ref{fig1}(a) and (c)).
These $\alpha$ and $\beta$ peaks were interpreted as arising
from the transitions from the occupied $J_{eff}$=1/2 lower Hubbard band (LHB)
and $J_{eff}$=3/2 band to the unoccupied $J_{eff}$=1/2 upper Hubbard band (UHB),
respectively, as shown in Fig.~\ref{fig1}(a).
In 214 system, both $\alpha$ and $\beta$ peaks are
clearly identified,
but, in 113 system, only the $\beta$ peak is identified
with the additional Drude contribution in the lower energy regime \cite{BJKim08,Liu13}.

Motivated by the idea of strain engineering for iridate systems,
we have investigated epitaxial-strain effects on the electronic structures
of two end members of Sr-iridates: Sr$_2$IrO$_4$ and SrIrO$_3$.
Our studies are based on the optical conductivity calculated by using the
\emph{ab initio} band method, which provides the direct comparison with experiments.
Also, cluster-based microscopic model calculations are employed
to do parameter-wise analysis of optical conductivity.
Note that a similar approach was applied to honeycomb iridate systems to
successfully explain key experimental findings \cite{Li15}.
Hybrid functional scheme with inclusion of the SOC term is employed,
and the results are analyzed and compared with various experimental strain studies
on Sr-iridates, especially, with
optical experiments \cite{Nichols13, Lupascu14, Liu13, Serrao13, Gruenewald14}.
We have found that the tensile strain on 214 system can effectively tune
the strengths of both electronic correlation and the SOC.
Strong interplay among the moment direction, the SOC, and the substrate strain in the
$J_{eff}=1/2$ ground state is reflected in the optical conductivities
as peak shifts or intensity changes of $\alpha$ and $\beta$ optical peaks.
On the other hand, in semimetallic 113 system, upon strain,
the $J_{eff}$=1/2 electronic structure is found to be rather fragile,
but low energy physics coming from narrow hole bands is found to be easily tunable.

\section{Results and discussions}

\subsection{Sr$_2$IrO$_4$}

\begin{table}[b]
\centering
\caption[]{
\textbf{Calculated Ir-O-Ir bond angle ($\theta$), Ir-O bond length ($d$) of 214 system
on different substrates.
}
Bulk results are also given for comparison.
}
\begin{tabular}{C{2.8cm}|C{1.4cm}C{1.4cm}C{1.4cm}|C{1.4cm}}
\hline\hline
                                 & LAO   & STO  & GSO  & bulk \\
\hline
    {Ir-O-Ir angle ($^{\circ}$)} & 152   & 156  & 159  & 157 \\
    {Ir-O length ({\AA})}      & 1.95  & 2.00 & 2.02 & 1.98 \\
\hline
\end{tabular}
\label{214_1}
\end{table}
Tensile strain increases both Ir-O-Ir angle ($\theta$) and Ir-O bond length ($d$) of IrO$_{6}$
octahedron, as shown in Fig.~\ref{fig1}(d).
The increases in $\theta$ and $d$ play mutually competing roles,
as the former enhances the bandwidth ($W$), while the latter localizes $5d$ electrons
to increase effective Coulomb correlation ($U$).
Recent optical experiment on 214 system showed the systematic shift of $\alpha$-peak
with enhanced broadening upon tensile strain \cite{Nichols13}.
This feature was explained by the enhancements of both $U$ and $W$,
which increase the separation of UHB and LHB and makes both bands
more dispersive, respectively.
As typical temperature-dependent behavior shows the enhancement
of one parameter with simultaneous suppression of the other,
the enhancements of both $U$ and $W$ are quite unusual \cite{Moon09}.

\begin{figure}[t]
\begin{center}
\includegraphics[angle=270,width=85mm]{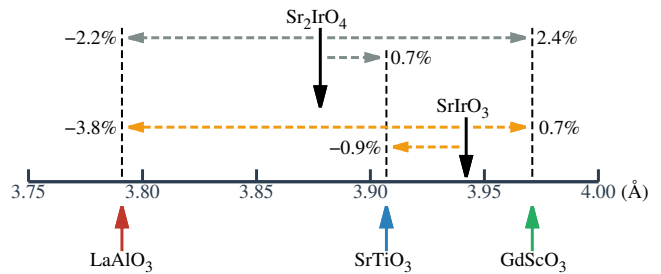}
\caption{
\textbf{Comparison of in-plane lattice parameter of Sr$_2$IrO$_4$ (214)
and SrIrO$_3$ (113) with various oxide substrates.
}
LaAlO$_3$ (LAO) and GdScO$_3$ (GSO) substrates yield compressive and tensile strains, respectively,
while the lattice mismatch would be minor for SrTiO$_3$ (STO).
In-plane lattices parameters for 214 and 113 systems are from
Ref. ~\cite{Crawford94} and Ref. ~\cite{Zhao08}, respectively,
where orthorhombic 113 system is converted to corresponding pseudo-cubic phase.
}
\label{fig2}
\end{center}
\end{figure}

To cover the epitaxial strain range of experimental reports,
we have chosen LaAlO$_3$ (LAO), SrTiO$_3$ (STO), and GdScO$_3$ (GSO) substrates.
As shown in Fig.~\ref{fig2},
LAO and GSO substrates yield compressive and tensile strains, respectively,
with +1.9\% and $-3.2$\% enhancements of $c/a$ ratio compared to bulk~\cite{Nichols13}.
In the case of the STO substrate, the lattice mismatch is small,
and so the corresponding $c/a$ ratio change is as small as $-0.6$\%.
Optimized $c/a$ ratio changes of LAO (+1.2\%), STO ($-2.1$\%), and GSO ($-5.3$\%)
cover well the experimental results.
Ir-O-Ir bond angle ($\theta$) and Ir-O bond length ($d$) of
corresponding 214 systems
are summarized in Table~\ref{214_1}.

Our calculation results for 214 films demonstrate more prominent role of $U$
than $W$ upon strain.
As shown in Table~\ref{214_m}, both spin and orbital magnetic moments
systematically increase,
as the substrate is changed from LAO to GSO,
along with corresponding shifts of optical peaks.
In accordance with our results,
recent resonant inelastic X-ray scattering (RIXS) experiment observed that
the most significant effect of substrate change is the variation of bond lengths,
which is manifested in the strengthening (weakening) of the magnetic interaction
of the 214 film upon compressive (tensile) strain~\cite{Lupascu14}.

\begin{figure}[t]
\begin{center}
\includegraphics[angle=270,width=85mm]{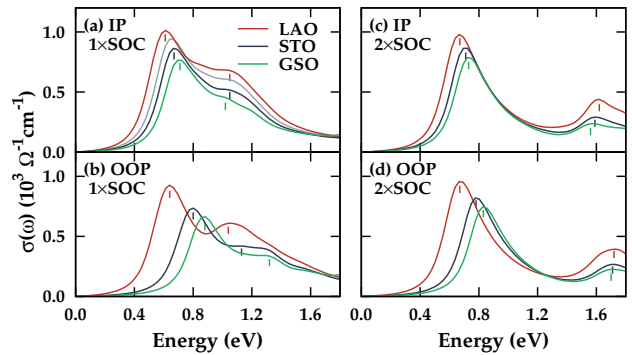}
\caption{
\textbf{Calculated optical conductivities $\sigma(\omega)$'s and DOSs for Sr$_2$IrO$_4$
on different substrates.
}
(a) IP-AFM ordering cases. Bulk $\sigma(\omega)$ (gray line) is also given, for comparison.
(b) Hypothetical OOP-AFM ordering cases.
(c),(d) Cases for doubled SOC strength (2$\times$SOC).
At the peak positions, small vertical lines are drawn for the guide to the eyes.
}
\label{fig3}
\end{center}
\end{figure}

To get the further insight of the role of the strain and to directly compare with the experiments,
we have calculated optical conductivity, $\sigma(\omega)$,
using the \emph{ab initio} band methods as described above.
Figure \ref{fig3}(a) presents the calculated $\sigma(\omega)$'s for 214 system
on different substrates.
$\sigma(\omega)$ for bulk is also presented for comparison.
Two-peak structure ($\alpha$ and $\beta$) is clearly manifested.
Note that, upon tensile strain, the position of $\alpha$ peak is shifted
to a higher energy side.
As schematically depicted in Fig.~\ref{fig1}(b) and (c), this feature is suggestive
of the enhancement of effective $U$,
which also agrees with the increase in the magnetic moment upon strain (Table~\ref{214_m}).
In contrast, the $\beta$ peaks are not affected much by the strain,
which suggests the different nature between $\alpha$ and $\beta$ peaks
(see Fig. \ref{fig1}(b) and (c)).
The peak positions of ($\alpha$ and $\beta$) are (0.61, 1.05), (0.67, 1.05), and (0.71, 1.02) eV
for LAO, STO, and GSO, respectively, which agree well with existing
experiment \cite{Nichols13}.

It is seen in Fig. \ref{fig3} that optical spectrum becomes broadened upon strain.
This strain-dependent broadening is interpreted as the increased itinerancy
due to change in the bond angle \cite{Nichols13}.
Despite the prominent role of $U$, as revealed by a shift of the $\alpha$-peak,
the broadening of optical spectrum would not be well described in our approach
due to lack of dynamical effect \cite{Zhang13}.
Thus, a possible explanation of broadening in Fig. \ref{fig3} is that
the tensile strain enhances the effective $U$, which reduces the coherency of the electrons.
Then, without much change in band width $W$, there occurs broadening of the peaks.
The difference between the temperature and the strain dependence of the optical conductivities
can be attributed to the altered coherency due to effective $U$ variation \cite{Moon09},
which is a subject of further studies.

Due to the two-dimensional (2D) character of the 214 system,
the overall optical responses are composed of in-plane characters
only ($\sigma_{xx}$ and $\sigma_{yy}$) \cite{suppl}.
Related densities of states (DOSs) information can be found in supplement materials~\cite{suppl}.

\begin{table}[b]
\centering
\caption[]{
\textbf{Calculated spin, orbital magnetic moments, their ratio,
and peak intensity ratio ($\mu_{S}$, $\mu_{O}$, $\mu_{O}/\mu_{S}$, and $I_{\beta}$/$I_{\alpha}$)
for 214 system on different substrates.
}
$I_{\beta}/I_{\alpha}$ here is defined by
$A_{\beta}{\epsilon_{\alpha}}/A_{\alpha}{\epsilon_{\beta}}$,
as described in Eq. \eqref{sigma} and below.
Bulk results are also given for comparison.
Unit of $\mu_{S}$ and $\mu_{O}$ is $\mu_{B}$/Ir.
}
\begin{tabular}{C{1.2cm}|C{1.0cm}|C{1.0cm}|C{1.0cm}C{1.0cm}C{1.0cm}|C{1.0cm}}
\hline\hline
                              &                    &                           & LAO   & STO  & GSO & bulk \\
\hline
\multirow{8}{*}{1$\times$SOC} &\multirow{4}{*}{IP} & $\mu_{S}$                 & 0.17 & 0.18 & 0.18 & 0.19 \\
                              &                    & $\mu_{O}$                 & 0.24 & 0.27 & 0.28 & 0.26 \\
                              &                    & $\mu_{O}/\mu_{S}$         & 1.43 & 1.52 & 1.58 & 1.40 \\
                              &                    & $I_{\beta}$/$I_{\alpha}$  & 1.31 & 1.08 & 1.04 & 1.22 \\
\cline{2-7}
                              &\multirow{4}{*}{OOP}& $\mu_{S}$                 & 0.26 & 0.36 & 0.40 & 0.29 \\
                              &                    & $\mu_{O}$                 & 0.32 & 0.38 & 0.41 & 0.34 \\
                              &                    & $\mu_{O}/\mu_{S}$         & 1.24 & 1.06 & 1.00 & 1.18 \\
                              &                    & $I_{\beta}$/$I_{\alpha}$  & 1.35 & 1.09 & 0.85 &  -   \\
\hline
\multirow{8}{*}{2$\times$SOC} &\multirow{4}{*}{IP} & $\mu_{S}$                 & 0.16 & 0.16 & 0.16 &      \\
                              &                    & $\mu_{O}$                 & 0.25 & 0.28 & 0.30 &  -   \\                             &                    & $\mu_{O}/\mu_{S}$         & 1.52 & 1.70 & 1.81 &      \\
                              &                    & $I_{\beta}$/$I_{\alpha}$  & 0.98 & 0.72 & 0.68 &  -   \\
\cline{2-7}
                              &\multirow{4}{*}{OOP}& $\mu_{S}$                 & 0.21 & 0.28 & 0.31 &      \\
                              &                    & $\mu_{O}$                 & 0.28 & 0.32 & 0.32 &  -   \\
                              &                    & $\mu_{O}/\mu_{S}$         & 1.33 & 1.14 & 1.05 &      \\
                              &                    & $I_{\beta}$/$I_{\alpha}$  & 0.97 & 0.66 & 0.63 &  -   \\
\hline
\end{tabular}
\label{214_m}
\end{table}

According to the previous studies on $J_{eff}$=1/2 systems,
the SOC and the tetragonality are crucial parameters
to stabilize the in-plane ordering of the system \cite{Jackeli09,Katukuri12,PLiu15}.
To investigate the roles of the SOC and the magnetic moment direction in determining
the strain-dependent electronic structure of the system,
we analyzed $\sigma(\omega)$'s (i) for different magnetic moment directions:
real in-plane (IP) and hypothetical out-of-plane (OOP) antiferromagnetic (AFM) orderings,
and (ii) for normal and enhanced SOC strengths.
As discussed below, the OOP configuration is related to the magnetic structure
of Sr$_3$Ir$_2$O$_7$ (327) system.
In Fig. \ref{fig3}(b), calculated $\sigma(\omega)$'s
for the OOP case are plotted.
Compared to the IP case, the OOP case shows quite different
response of the electronic structure to the substrate strain.
The overall shifts are very large for the OOP case.
As the substrate changes from LAO to GSO, $\alpha$ peak positions change by 0.10 eV
and 0.24 eV for the IP and the OOP, respectively,
while $\beta$ peak positions change by $-0.03$ eV and 0.28 eV for the IP and the OOP,
respectively. Namely, when the 214 system has the IP-AFM ordering,
the electronic structure is rather robust
against the epitaxial strain, whereas, when the system has the OOP-AFM ordering,
the overall electronic structure becomes more susceptible to the strain.
In fact, Boseggia \emph{et al.}\cite{Boseggia13}
linked the IP magnetic ordering in 214 to the $J_{eff}$=1/2 electronic structure,
on the basis of
its insensitiveness to the structural distortion,
which is in agreement with our calculations.

When the SOC strength of the system is doubled (2$\times$SOC),
the most pronounced effect is the large shift-down in energy of $J_{eff}=3/2$ state,
as schematically plotted in Fig. \ref{fig1}(b) and (c),
which is reflected by the huge shift-up of the $\beta$ peak in Fig. \ref{fig3}(c).
Another notable change is the reduction in the relative intensity
of $\alpha$ and $\beta$ peak ($I_{\beta}$/$I_{\alpha}$).
As each substrate case has different $\omega_{\beta}/\omega_{\alpha}$ value
(1.72, 1.57, and 1.44 for LAO, STO and GSO (for 1$\times$SOC IP case)) and
as there is $1/{\omega}$ dependence in the optical conductivity,
the intensity is not to be defined by the height of each peak.
We have quantitatively analyzed the intensities within a two-peak picture,
taking into account the $1/{\omega}$ dependence of the optical conductivity curve,
and fitted the data with following Lorentzian-type equation:
 \begin{equation}
\sigma(\omega) = \frac{A_{\alpha}} {\omega_{\alpha}} \frac{\pi^{-1} \epsilon_{\alpha}}
  { (\omega-\omega_{\alpha})^2 + {\epsilon_{\alpha}}^2\ } +
\frac{A_{\beta}} {\omega_{\beta}} \frac{\pi^{-1} \epsilon_{\beta}}
  { (\omega-\omega_{\beta})^2 + {\epsilon_{\beta}}^2\ },
\label{sigma}
\end{equation}
where we can define peak intensity at each frequency position
as $I_{\alpha}=A_{\alpha}{\pi^{-1}}/{\epsilon_{\alpha}}$
or $I_{\beta}=A_{\beta}{\pi^{-1}}/{\epsilon_{\beta}}$.

As Kim \emph{et. al.} \cite{BHKim12} have shown,
the $\beta$ peak, that is thought to arise
from transition from low-lying $J_{eff}$=3/2 band
to $J_{eff}$=1/2 UHB in a simple picture,
has in fact large $J_{eff}$=1/2 LHB contributions.
With increasing the SOC parameter,
$J_{eff}$=1/2 and $J_{eff}$=3/2 bands are decoupled,
and $I_{\beta}$/$I_{\alpha}$ is diminished
because of the reduction of $J_{eff}$=1/2 contribution to $\beta$ peak.
Namely, the effective increase of the SOC strength can be identified
as the decrease of $I_{\beta}$/$I_{\alpha}$.
We can clearly see the reduction of $I_{\beta}$
with respect to $I_{\alpha}$ for 2$\times$SOC cases in Fig. \ref{fig3}(c) and (d),
regardless of moment directions and substrate types (see Table ~\ref{214_m}).

Surprisingly, $I_{\beta}$/$I_{\alpha}$ ratio is found to decrease systematically upon strain,
as shown in Table ~\ref{214_m} for different substrate strain cases.
This feature suggests that the tensile strain acts similarly to the increased SOC strength.
The ratio of orbital and spin magnetic moment ($\mu_{O}/\mu_{S}$) also shows similar trend.
As the tensile strain is applied, the $\mu_{O}/\mu_{S}$ value increases
and approaches to 2 (see Table~\ref{214_m}),
which corresponds to a value for the ideal $J_{eff}$=1/2 state
of strong SOC limit.
The $\beta$ peak shift, which occurs for increased SOC strength (2$\times$SOC),
has been observed in the experiment \cite{Nichols13},
even though it is not identified within our studied substrate-strain range.
This feature indicates that the SOC can be enhanced effectively
by means of the tensile strain.
However, according to the atomic microscopic model,
the strain-dependent hopping parameter is also found to produce similar optical behavior
for a fixed SOC strength.
Thus the overall optical behaviors are expected to come from combined effects of
both the SOC strength and hopping parameters.
For the OOP-AFM case, upon tensile strain,
similar reduction of $I_{\beta}$/$I_{\alpha}$ is obtained,
but $\mu_{O}/\mu_{S}$ decreases as opposed to the IP case (see Table~\ref{214_m}).
This feature occurs due to the eventual breakdown of $J_{eff}$=1/2
electronic state rather than the increase in the SOC strength.

Table ~\ref{214_2} provides the  band gap dependence on the magnetic moment direction
in 214 system.
Considering that the ideal $J_{eff}$=1/2 picture is validated
in the insulating limit,
the overall increasing behavior of the band gap upon strain
is quite reasonable.

\begin{table}[b]
\centering
\caption{
\textbf{Band gaps (in eV) of 214 system on different substrates,
depending on the SOC strength and magnetic moment direction.
}
Bulk results are also given for comparison.
}
\begin{tabular}{C{1.0cm}C{2.0cm}|C{1.3cm}C{1.3cm}C{1.3cm}|C{1.3cm}}
\hline\hline
 & moment direction & LAO   & STO  & GSO & bulk \\
\hline
  \multirow{2}{*}{1$\times$SOC} & IP & 0.14 & 0.28 & 0.34 & 0.21 \\
                                & OOP  & 0.19 & 0.44 & 0.57 & 0.28 \\
\hline
  \multirow{2}{*}{2$\times$SOC} & IP & 0.28 & 0.40 & 0.46 & - \\
                                & OOP  & 0.47 & 0.49 & 0.58 & - \\
\hline
\end{tabular}
\label{214_2}
\end{table}

\begin{figure}[t]
\begin{center}
\includegraphics[angle=0,width=85mm]{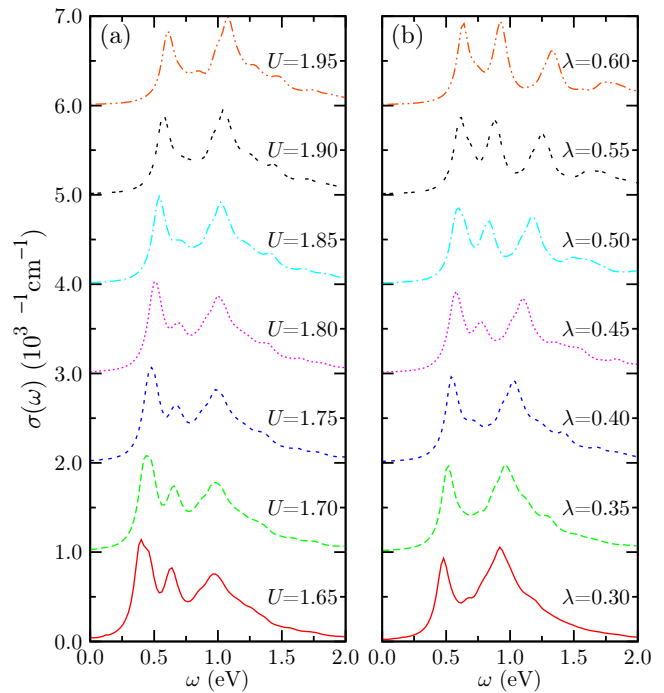}
\caption{
\textbf{Optical conductivities obtained by four-site cluster multiplet calculations for
various on-site Coulomb strengths ($U$) and  spin-orbit strengths ($\lambda$).
}
Other parameters are the same as in Ref.~\cite{BHKim12}.
}
\label{fig4}
\end{center}
\end{figure}

 To confirm the enhanced $U$ and SOC behaviors upon strain,
we obtained $\sigma(\omega)$ using
the microscopic model calculations with varying physical parameters.
 Figure~\ref{fig4}(a) and (b) presents $\sigma(\omega)$'s with respect to $U$ and $\lambda$, respectively.
Dominant optical spectra are attributed to the electron-hole (e-h) excitations
in the vicinity of the Mott gap.
With increasing $U$,
the optical peaks shift up due to the enhancement of Mott gap.
In addition, the shape of optical spectrum varies depending on $U$ values.
The change from three-peak to two-peak structure is observed.
Interesting finding is that the middle-peak is depleted
when the shape of optical spectrum changes.
It is expected to occur due to
the Fano-type coupling between the spin-orbit (SO) exciton and e-h excitation of
$J_{eff}=1/2$ band~\cite{BHKim12}.
Whether three-peak structure really appears in $\sigma(\omega)$ of iridate is not so certain,
because the four-site cluster we have considered in Fig.~\ref{fig4} may not be sufficient
to describe full kinetics of lattice.
However, it is legitimate to infer that some optical spectral-weight transfer to
higher peak ($\beta$ peak) occurs with increasing $U$,
which corresponds to tensile strain behavior.

When the SOC increases, the splitting between $J_{eff}=1/2$ and $3/2$ bands increases
and the Mott gap is slightly enhanced.
These features are well reflected in the optical conductivity shown in Fig.~\ref{fig4}(b).
The lowest energy peak becomes slightly higher and the highest energy peak shifts up somehow,
when $\lambda$ increases.
As in the case of the  weak $U$ ($<$1.8 eV),
a three-peak structure appears
for large $\lambda$ ($>$ 0.45 eV).
It happens because the Fano-type coupling is weakened
as the excitation energy
of the SO exciton becomes higher than that of the e-h excitations.
Because some spectral weights depleted for small $\lambda$ are recovered,
the spectral weight near the $\beta$ peak diminishes when $\lambda$ becomes larger.
$I_{\beta}$/$I_{\alpha}$ behavior shows the reduction upon increasing $\lambda$, which
also mimics the tensile strain effect in the \emph{ab initio}-based optical data,
suggesting the effective increase of SOC strength.

Note that, in the current model approach,
$I_{\beta}$/$I_{\alpha}$ is also affected by the change in
the hopping parameters due to substrate strain,
namely, the enhanced hopping between $J_{eff}=1/2$ bands and
the reduced hopping between $J_{eff}=1/2$ and $J_{eff}=3/2$ bands,
under the tensile strain.
Thus, regarding the peak intensities,
the enhanced optical spectral weight of e-h excitation of $J_{eff}=1/2$
is expected to yield similar effect to the enhanced SOC strength.

In general, care should be taken for applying low-energy atomic model to itinerant 5$d$ system.
Since the intensity of $\sigma(\omega)$ in the model approach
is obtained by the sum of possible four spectral weights from $d^4-d^6$
multiplet configurations \cite{BHKim12}, the analysis of the each spectral weight
upon parameter change is possible.
In the \emph{ab initio} methods however, the strain-dependent change in $I_{\beta}$/$I_{\alpha}$
can be the result of cooperative changes in many physical parameters, not solely from SOC strength.
As we have seen in the IP and OOP cases, the additional information on $\mu_{O}/\mu_{S}$ change
is necessary to conclude that the primary tuning parameter in the IP case is
the SOC strength, while it is not in the OOP case.

The opposite behavior of $\mu_{O}/\mu_{S}$ for IP and OOP can also be understood
in terms of a simple atomic picture.
For a state close to ideal $J_{eff}$=1/2 state, $\mu_{O}/\mu_{S}$ can be expressed as
\begin{eqnarray}
 {\mu_{O}/\mu_{S}}  & =   \frac{4}{\sqrt{\delta^2-2\delta+9}+\delta-1}
	\simeq 2 (1-\frac{1}{3}\delta)~\text{: IP} \label{mu_IP}, \\
 {\mu_{O}/\mu_{S}}  & =   \frac{4(1-\delta)^{-1}}{\sqrt{\delta^2-2\delta+9}+\delta-1}
	\simeq 2 (1+\frac{2}{3}\delta) ~\text{: OOP}, \label{mu_OOP}
\end{eqnarray}
where $\delta = \frac{2\Delta}{\lambda}$ ($\lambda$: SOC strength)
represents small deviation from the ideal cubic case
due to tetragonal crystal field splitting ($\Delta$)
(see supplement materials for the derivation)~\cite{suppl}.

Considering itinerant character of 5$d$ system,
atomic model may not access the full description of the system,
but the strain dependency is expected to be well-described.
As $\delta$ goes more negative upon tensile strain,
the IP (OOP) case shows clear increase (decrease) in $\mu_{O}/\mu_{S}$.
The more rapid decrease for the OOP case agrees well
with the tendency shown in Eq. \eqref{mu_IP} and \eqref{mu_OOP} (see Table~\ref{214_m}).

The different strain dependence between 214 and Sr$_3$Ir$_2$O$_7$ (327) system
are also expected to come from the different magnetic moment directions,
as the former and the latter have IP and OOP-AFM orderings, respectively.
The strain dependence of 327 system resembles the hypothetical OOP-AFM phase
of 214 system \cite{Zhang13},
which suggests that the response of electronic structure upon strain is
more related to the moment direction than to the dimensionality of the Sr-iridates.
The tensile strain can effectively change the $J_{eff}$=1/2 nature of the system
through the change of the moment direction as well as
the change in the electronic correlation \cite{PLiu15}.
In conjunction with recent analysis on resonant X-ray scattering of iridate systems,
we corroborate that the moment direction plays a role of another degree of freedom
that can be tuned using the substrate engineering,
especially, for a system with many competing energy scales \cite{Sala14_2,anisotropy}.

The 2$\times$SOC cases show overall similar strain trends,
but with more robustness of the electronic structures.
As can be seen by increased $\mu_{O}/\mu_{S}$
along with reduced $I_{\beta}$/$I_{\alpha}$ (Table~\ref{214_m}),
the electronic structure for the 2$\times$SOC becomes closer
to that of $J_{eff}$=1/2 state, which is reflected by highly reduced
optical peak shifts upon external strain in Fig.\ref{fig3}(c).

 According to Jackeli \emph{et al.},\cite{Jackeli09}
the magnetic moment direction of the $J_{eff}$=1/2 system
can be changed by changing tetragonality.
In our studied substrate-range, however,
the magnetic structure of the system remains IP-AFM,
which is more stable than OOP-AFM by ~100 meV/{f.u.}.
The 2$\times$SOC cases show even larger energy difference between IP and OOP-AFM,
which indicates the strong interconnection between the $J_{eff}$=1/2 electronic structure
and the magnetic moment direction of the system.
According to recent study \cite{PLiu15},
much larger tetragonality change is needed for the change of magnetic moment direction.

Before we move on, we want to comment on the possible magneto-electric effect.
As the tensile strain increases, the overall electronic structures of 214 system
between the IP and OOP cases become progressively distinct,
which is revealed by the differences in $I_{\beta}$/$I_{\alpha}$, $\mu_{O}/\mu_{S}$,
and the optical conductivity shapes for different substrates
(Table~\ref{214_m} and Fig. \ref{fig3}).
Different electronic structures between IP and OOP moment directions
can be used to generate strong magneto-electric effect,
especially for strained system, as such by applying the strong magnetic field.
Namely, the control of the electronic nature, such as optical gap,
would be feasible by employing strained iridate systems.

\subsection{SrIrO$_3$}

 Differently from 214 system, 113 system is known as a correlated metal
with semimetallic character,
being located at the boundary of the magnetic metal
and magnetic insulator in the phase diagram \cite{Moon08,Zhang13,Zeb12}.
We have found that 113 system is
to be a paramagnetic metal for all studied substrate-strain range.
In 113 system, the response of the electronic structure to the epitaxial strain
is expected to be reduced
with respect to the case in 214 system,
due to the 3D connectivity of the IrO$_6$ octahedra.
As shown in Fig.~\ref{fig1}(d),
the in-plane strain effects are compensated
by the change in apical connectivity of the IrO$_6$ network,
which can be seen in the bond length and bond angle variation upon strain Table~\ref{113}.

\begin{table}[b]
\centering
\caption{
\textbf{Calculated Ir-O-Ir bond angle ($\theta$), Ir-O bond length ($d$)
of 113 system on different substrates.}
}
\begin{tabular}{C{2.3cm}C{1.5cm}|C{1.4cm}C{1.4cm}C{1.4cm}}
\hline\hline
 &  & LAO   & STO  & GSO \\
\hline
  \multirow{2}{*}{Ir-O-Ir angle ($^{\circ}$)} & apical & 160 & 155 & 152 \\
                                               & in-plane  & 150 & 153 & 155 \\
\hline
  \multirow{2}{*}{Ir-O length ({\AA})} & apical & 2.06 & 2.02 & 1.99 \\
                                        & in-plane  & 1.96 & 2.01 & 2.04 \\
\hline
\end{tabular}
\label{113}
\end{table}

\begin{figure}[t]
\begin{center}
\includegraphics[angle=270,width=85mm]{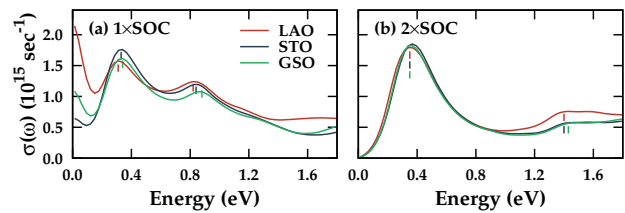}
\caption{
\textbf{Calculated optical conductivities for SrIrO$_3$ (113).}
Cases with (a) normal (1$\times$SOC) and (b) doubled SOC term (2$\times$SOC) on different substrates.
At the peak positions, small vertical lines are drawn for the guide to the eyes.
}
\label{fig5}
\end{center}
\end{figure}

 Optical experiment for 113 system has shown that the $\beta$ peak position is shifted
to a higher energy side as the tensile strain is applied,
while the $\alpha$ peak is not clearly identified \cite{Liu13}.
We also obtained the shift of $\beta$ peak by 0.06 eV from LAO to GSO substrate
(Fig. \ref{fig5}(a)). The $\alpha$ peak, which has not been identified in experiment,
appears in our calculation
due to the incapability of describing the dynamical correlation effect \cite{Kotliar04,Zhang13}.
Since the 113 system is weakly correlated,
careful change of relative $W$ and $U$ parameters using substrate strain
would produce the correlated three-peak structure in the DOS
to locate the $\alpha$ peak in the vicinity of the Drude part.
Due to 3D connectivity of 113 system,
$\alpha$ peak shifts are highly suppressed (0.03 eV shift from LAO to GSO case)
with respect to the case in 214 system.
Note in Fig. \ref{fig5}(a) that, upon tensile strain,
the systematic separation of $\alpha$ and $\beta$ peaks occurs
with the reduction of the $\beta$ peak intensity,
as observed in 214 system.
According to recent experiments for 113 films,
the position of the $\beta$ peak under the small compressive strain
shows only a little shift \cite{Liu13,Gruenewald14}.
Considering that our calculation covers wider range of strain,
further experiments with various substrates
are demanded to get more information on the substrate effects.
Also, recent finding of enhanced scattering
for the compressive strain case, which was ascribed to the disorder effect
rather than to the correlation effect,\cite{Biswas14}
can also be justified by examining the $\alpha$ peak shift upon substrate strain.

 For the 2$\times$SOC case in Fig. \ref{fig5}(b), much larger shift-up of
$\beta$ peaks is shown, as in 214 system.
Again, the enhanced $J_{eff}=1/2$ ground state of the system is well-described
with highly reduced $I_{\alpha}$/$I_{\beta}$
and with more insulating nature in the DOS \cite{suppl}.
Combined with 3D nature of the system,
the enhanced SOC highly stabilizes electronic structure against strain,
which is evident from almost locking of both optical peaks in Fig. \ref{fig5}(b).
All the substrate cases for 2$\times$SOC are almost insulating
with no Drude contribution
in $\sigma(\omega)$ in agreement with the reported \emph{ab initio} phase diagram \cite{Zeb12}.

\begin{table}[b]
\centering
\caption{$I_{\beta}/I_{\alpha}$ of 113 systems on different substrates.}
\begin{tabular}{C{1.5cm}|C{1.5cm}C{1.5cm}C{1.5cm}}
\hline\hline
 & LAO   & STO  & GSO\\
\hline
  {1$\times$SOC} &  2.59 & 1.96 & 1.89\\
\hline
  {2$\times$SOC} &  1.21 & 0.90 & 0.92\\
\hline
\end{tabular}
\label{113_2}
\end{table}

In the case of 214 system, the IP-AF ordering was essential
for the effective tuning of SOC, while the OOP-AF shows the break down of
$J_{eff}=1/2$ picture upon strain.
For nonmagnetic 113 system,
the reduction of $I_{\beta}$/$I_{\alpha}$ cannot be claimed to be due to
enhancement of effective SOC (see Table~\ref{113}).
According to recent reports, the ground state of 113 has
large deviation from $J_{eff}=1/2$ state
and the mixing of $J_{eff}=1/2$ and $J_{eff}=3/2$ is found to be significant
with the entrance of octahedral rotations,
which is in sharp contrast to layered 214 system \cite{Nie15,Liu15}.
Since the substrate strain directly changes the octahedral rotations,
we can deduce that the shift and reduction of $\beta$ peak in 113 system are
due to the deviation of ground state from $J_{eff}=1/2$ state,
and $I_{\beta}$/$I_{\alpha}$ reduction is due to enhanced optical spectrum weight
of $J_{eff}=1/2$ e-h excitation,
which is totally different from the case in 214 system.

\begin{figure}[t]
\begin{center}
\includegraphics[angle=0,width=85mm]{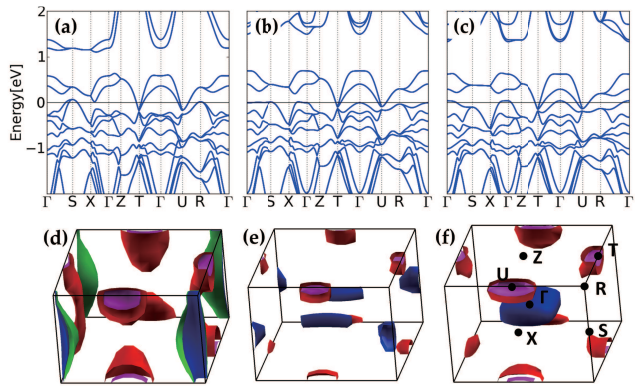}
\caption{(Color online)
\textbf{Band structure and Fermi surfaces of 113 system on different substrates
}
(a)-(c) band structure for LAO, STO, and GSO substrates.
(d)-(f) Corresponding Fermi surfaces.
}
\label{fig6}
\end{center}
\end{figure}

 Finally, we want to discuss the low-energy electronic structure of 113 system upon strain.
Even though the strain dependency is highly reduced
with respect to 214 system due to the dimensionality change,
the narrow-band semimetallic nature
of 113 system near the Fermi level ($E_F$) makes the system very tunable
upon small change of external parameters in low-energy scales.
As shown in Fig. \ref{fig6}(a)-(c), overall band structures of 113 system
on different substrates are similar to that of bulk system \cite{Zeb12},
but there are a few points to be pointed out.
First, we found that 113 system on STO has almost cubic electronic structure,
which can be recognized by the highest $e_g$ band location near 2 eV above $E_F$.
The tetragonal crystal field in the presence of the substrate strain
lifts the degeneracy of $e_g$ states
with lowering one out of two $e_g$ states
($z^2$ for LAO and $x^2-y^2$ for GSO) toward $E_F$.
Second, while the electron pockets are retained at $\mathbf{k}=T$ and $U$,
hole pockets emerge at different $\mathbf{k}$'s depending on the strain, {\it i.e.}
at $\mathbf{k}=S$ and $R$ for LAO and near $\mathbf{k}=\Gamma$ for GSO.
For the STO case, the morphologies of hole pockets are in-between LAO and GSO cases,
with very narrow band character near $\Gamma$-$S$ and $R$-$\Gamma$,
which enables easy tune upon the epitaxial strain.
The Fermi surface topology also changes accordingly, as shown in Fig. \ref{fig6}(d)-(f).

 In relation to recent experiments,
heavier effective mass of hole carriers than electron carriers \cite{Liu13,Nie15}
can be identified from the band structure of STO substrate case (Fig. \ref{fig6}(b)).
More symmetric electron-hole band structure
for tensile strain case is also consistent with transport measurement \cite{Liu13}.
Under compressive strain, electron pockets at $U$ and $T$, and hole pocket
at $R$ are formed, as shown in Fig. \ref{fig6}(a) and (d),
which are in good agreement with angle-resolved photoemission
spectroscopy (ARPES)~\cite{Liu15}.

In view of small band renormalization factor of 1.25
from ARPES experiment for 113 system \cite{Liu15},
our results successfully explain the low-energy electronic structure
for both compressive and tensile strain cases,
and suggest further possibility of manipulating the strain engineering.
Inconsistency of simple tight-binding model with ARPES
may come from the highly susceptible low-energy electronic structure
of the system \cite{Zhang15,Liu15}.
Suggested Dirac-cone-like nodes at $U$ and $T$ from tight-binding calculation
were not detected in the recent ARPES measurement \cite{Carter12,Liu15}.
Our band structure shows the protected node at $T$ upon strain
but no Dirac-node
at $U$, which needs confirmation by further experimental studies.
As the band structure of 113 system depends highly on the $U$ value,
which is interconnected to the SOC strength,
a proper estimation of electronic correlation $U$ value is crucial
from the theory side \cite{Carter12,HSKim15}. Also, recent study on 113
film demonstrated that the breaking of the crystal symmetry upon strain can lift the Dirac
node of 113 system \cite{JLiu15},
which reflects that the electronic structure of the
system is highly tunable upon systematic epitaxial strain.

\section{Conclusion}
We have analyzed the substrate strain effects in Sr-iridate systems,
employing both the \emph{ab initio} optical conductivity calculation
and the microscopic model approach.
By analyzing optical peak positions and relative intensities
along with obtained magnetic moment,
we have found that, in layered 214 system, tensile strain can effectively tune
the electronic correlation strength $U$ as well as the SOC strength.
The robustness of the $J_{eff}=1/2$ electronic structure,
which is found to be highly correlated with the magnetic moment direction of the system,
can also be controlled
by employing the substrate strain effect.
On the other hand, in 113 system,
tensile strain easily breaks the overall $J_{eff}=1/2$ ground state,
and band topology shows highly tunable hole character in the vicinity of $E_F$.
Our systematic study demonstrate that the strain engineering for iridate systems,
in which various energy scales compete,
provides an additional degree of freedom of tunable parameters, $U$ and SOC,
as shown as peak and weight change of the optical conductivity,
which can offer new dimensions on top of the current epitaxial strain studies,
especially, when combined with very recent studies based on superlattice structures \cite{Matsuno15}.


%

\section{Methods}

\subsection{\emph{Ab initio} calculation}
%
We have performed electronic structure calculations,
employing the full-potential linearized augmented plane wave (FLAPW) band
method\cite{Weinert82,Jansen84} implemented in WIEN2k package \cite{Wien2k}.
For the exchange-correlation energy functional, we used the local density approximation (LDA),
which has been generally employed for 5$d$ systems.
To treat the correlation in functional level, we employed the hybrid-functional \cite{Moreira02,Tran06}, which is given by
\begin{equation}
E^{hyb}_{xc}[\rho]=E^{LDA}_{xc}[\rho]+\gamma(E^{HF}_{x} [\Psi_{corr}]-E^{LDA}_{x}[\rho_{corr}]).
\end{equation}
Here $\Psi_{corr}$ and $\rho_{corr}$ correspond to the Kohn-Sham wave function
and the electron density of correlated electrons, respectively.
The exchange-correlation energy functional is constructed with the fraction ($\gamma$)
of the Hartree-Fock (HF) exchange energy, replacing the LDA correspondence
for correlated electrons (5$d$-electrons in the present case).
This functional form is the LDA correspondence of so-called PBE0 \cite{Ernzerhof99,Adamo99}.
Compared to the normally employed LDA$+U$ method, the hybrid-functional approach
can treat the correlation effects of different systems in a consistent way
and the non-local exchange energy can be included in the HF term.
The hybrid-functional scheme has been employed
for numerous transition-metal (TM) perovskites,
from 3$d$ to 5$d$ systems, and is thought to be one of the best computational schemes \cite{Franchini14}.
Especially for more itinerant 5$d$ systems, recent calculation found hybrid functional scheme
successfully described the electronic structures and magnetic properties \cite{Gangopadhyay15}.
The important SOC term is included in the second variational scheme.

\begin{table}[b]
\centering
\caption{
\textbf{Band gap dependence on the size of mixing parameter $\gamma$ for LDA and PBEsol functionals.
}
Calculations were done for the experimental bulk Sr$_2$IrO$_4$ (214) system.
}
\begin{tabular}{C{3.0cm}|C{1.5cm}C{1.5cm}C{1.5cm}}
\hline\hline
functional    & $\gamma$=0.15   & $\gamma$=0.20  & $\gamma$=0.25 \\
\hline
    LDA (eV) & metal & 0.21 & 0.37\\
    PBEsol (eV) & metal & 0.26 & 0.41\\
\hline
\end{tabular}
\label{gamma}
\end{table}

To determine the proper $\gamma$ parameter, we performed the calculations
on bulk Sr$_2$IrO$_4$ (214) system
with various $\gamma$ values, using both the LDA and PBEsol functionals.
As shown in Table~\ref{gamma}, both functionals show similar results of increasing gap size with $\gamma$.
Considering the observed optical gap size of around 0.4 eV, $\gamma$ value
in-between 0.20 and 0.25 looks appropriate. In the present study, we chose the LDA functional with $\gamma$=0.20
to fit the observed optical peak positions. However, the overall strain dependency is expected to be similar
for various $\gamma$ values.
Our choice of $\gamma$=0.20 is somewhat smaller than the often-used typical value of $\gamma$=0.25.
But the systematic studies for the perovskite systems showed that the smaller value of $\gamma$
produces much better results \cite{Franchini14}.

The substrate strain effects were taken into account by fixing
in-plane(IP) lattice parameters of 214 and SrIrO$_3$ (113) systems
to those of the substrates: LaAlO$_3$  (LAO), SrTiO$_3$ (STO), and GdScO$_3$ (GSO).
Since the relevant optical experiments were performed not on ultrathin films,
we did not consider the substrate materials explicitly.
We assumed the collinear magnetic structures for both IP and out-of-plane (OOP) cases
based on the fact that
the IP ferromagnetic (FM) component due to the canted antiferromagnetic (AFM) structure
is substantially weakened for the film case \cite{Miao14}.

We optimized $c/a$ ratio first with fixed $a$, which determines tetragonality of the system,
and then performed the internal relaxations for given volume of every systems
with force criteria of 1.0 mRy/a.u. within the LDA limit.
With obtained structures, we performed the hybrid-functional calculations with the inclusion of the SOC term.
In a system where the SOC plays a dominant role, inclusion of non-diagonal parts
of the spin density matrices are found to be crucial.
Especially, for iridates, inclusion of only diagonal parts does not describes the energy gap
and magnetic moments of the system \cite{MJKim15},
which even changes the energetics of the 214 system.
Without non-diagonal parts, the magnetic moment direction of the system is found to be OOP,
 which is corrected only after the inclusion of the full matrix elements.
In addition to hybrid functional parts, we included non-diagonal elements
of density matrices corresponding $U$=2 eV in generating orbital potentials,
for the description of weakly correlated Ir 5$d$ electrons.
The valence wave functions inside the muffin-tin spheres were expanded with spherical harmonics
up to $l_{max}$=10. The wave function in the interstitial region was expanded with plane waves
up to $K_{max}=7.0/R_{MT}$, where $R_{MT}$ is the smallest muffin-tin sphere radius.
$R_{MT}$ were set as 2.3, 2.1, and 1.5 a.u. for Sr, Ir, and O, respectively.
The charge density was expanded with plane waves up to $G_{max}$=12 (a.u.)$^{-1}$.
We have used 1000 $\mathbf{k}$ points inside the first Brillouin zone for both 214 and 113 systems.

Optical conductivity is calculated with the WIEN2k optical package
with much denser $\mathbf{k}$ points up to 3000 \cite{Draxl06}.
The dielectric function is calculated using the following expression:

\begin{eqnarray}
\label {eq:diel}
  \mathrm{Im} \epsilon_{\alpha\beta}(\omega) &=&
  \frac{4\pi e^2}{m^2\omega^2} \sum \limits_{n,n'} \int d\mathbf{k}\langle n_\mathbf{k}|p^{\alpha}|n'_\mathbf{k}\rangle
	\langle n'_\mathbf{k}|p^{\beta}|n_\mathbf{k}\rangle  \nonumber \\
	& & \times \delta (\varepsilon_{n_\mathbf{k}}-\varepsilon_{n'_\mathbf{k}}-\omega),
\end{eqnarray}
where the transition matrix of the momentum operator $p^{\alpha}$ between Kohn-Sham states represented by band index $n$ and crystal momentum $\mathbf{k}$ with energy $\varepsilon_{n_\mathbf{k}}$ is evaluated and summed. The optical conductivity can be obtained from the Kramer-Kronig transformation,
\begin{equation}
\label {eq:op_cond}
 \mathrm{Re} \sigma_{\alpha\beta}(\omega) = \frac{\omega}{4\pi} \mathrm{Im} \epsilon_{\alpha\beta}(\omega).
\end{equation}
For metallic 113 system, the Drude contribution of the following form is considered,
\begin{equation}
\label {eq:op_drude}
 \sigma_{D}(\omega) = \frac{\Gamma{\omega_p}^2}{4\pi(\omega^2 + \Gamma^2)},
\end{equation}
 where $\Gamma$ is lifetime broadening and $\omega_p$ is the plasma frequency given by
\begin{eqnarray}
\label {eq:plasma}
  {\omega^2_{p;\alpha\beta}} &=&
  \frac{\hbar^2 e^2}{\pi m^2} \sum \limits_{n} \int d\mathbf{k}\langle n_\mathbf{k}|p^{\alpha}|n_\mathbf{k}\rangle \langle n_\mathbf{k}|p^{\beta}|n_\mathbf{k}\rangle \nonumber \\
   & & \times \delta (\varepsilon_{n_\mathbf{k}}-\varepsilon_{F}).
\end{eqnarray}

We adopted the Gaussian broadening parameters of the interband transition of the value of 0.10 eV.
For the Drude contribution in 113 system, we set 0.10 eV for broadening parameter
to describe metallic and semimetallic characters of the systems.

\subsection{Microscopic model calculations}
Based on four-site cluster calculation
including all possible Ir multiplets
among $d^5-d^5-d^5-d^5$ and $d^4-d^6-d^5-d^5$ charge configurations,
we have solved the effective magnetic Hamiltonian with the exact diagonalization (ED) method.
$\sigma(\omega)$ is obtained from the following expression:
\begin{equation}
\label {eq:sigma_omega}
\sigma(\omega) = \pi \upsilon \frac{1-e^{-\beta \omega}}{\omega}
\sum \limits_{n<m} p_n |\langle \psi_m|\hat{J}_c|\psi_n \rangle|^2 \delta(\omega+E_n-E_m) ,
\end{equation}
where $\upsilon$ is the volume per Ir site, $p_n$ is probability density
 of eigenstate $|\psi_n \rangle$, and $\hat{J}_c$ is current operator.
See Ref.~\cite{BHKim12} for the details of calculation method and parameters.

\acknowledgments

 We would like to thank Y. H. Jeong and S. S. A. Seo for helpful discussions.
 This work was supported by the NRF (Grant No. 2011-0025237),
the National Creative Initiative (Grant No. 2009-0081576),
Max-Plank POSTECH/KOREA Research Initiative (Grant No. KR 2011-0031558),
and by the KISTI supercomputing center (Grant No. KSC-2014-C3-040). \\ \\


\pagebreak
\widetext
\begin{center}
\textbf{{\it Supplementary Material:}\\
    Substrate-tuning of correlated spin-orbit oxides}
\end{center}

\renewcommand{\thefigure}{S\arabic{figure}}
\renewcommand{\thetable}{S\arabic{table}}
\renewcommand{\theequation}{S\arabic{equation}}
\renewcommand{\bibnumfmt}[1]{[S#1]}
\renewcommand{\citenumfont}[1]{S#1}

\setcounter{equation}{0}
\setcounter{figure}{0}
\setcounter{table}{0}
\setcounter{page}{1}
\makeatletter

\section{component-wise optical conductivity}

\begin{figure}[h!]
\includegraphics[angle=270,width=85mm]{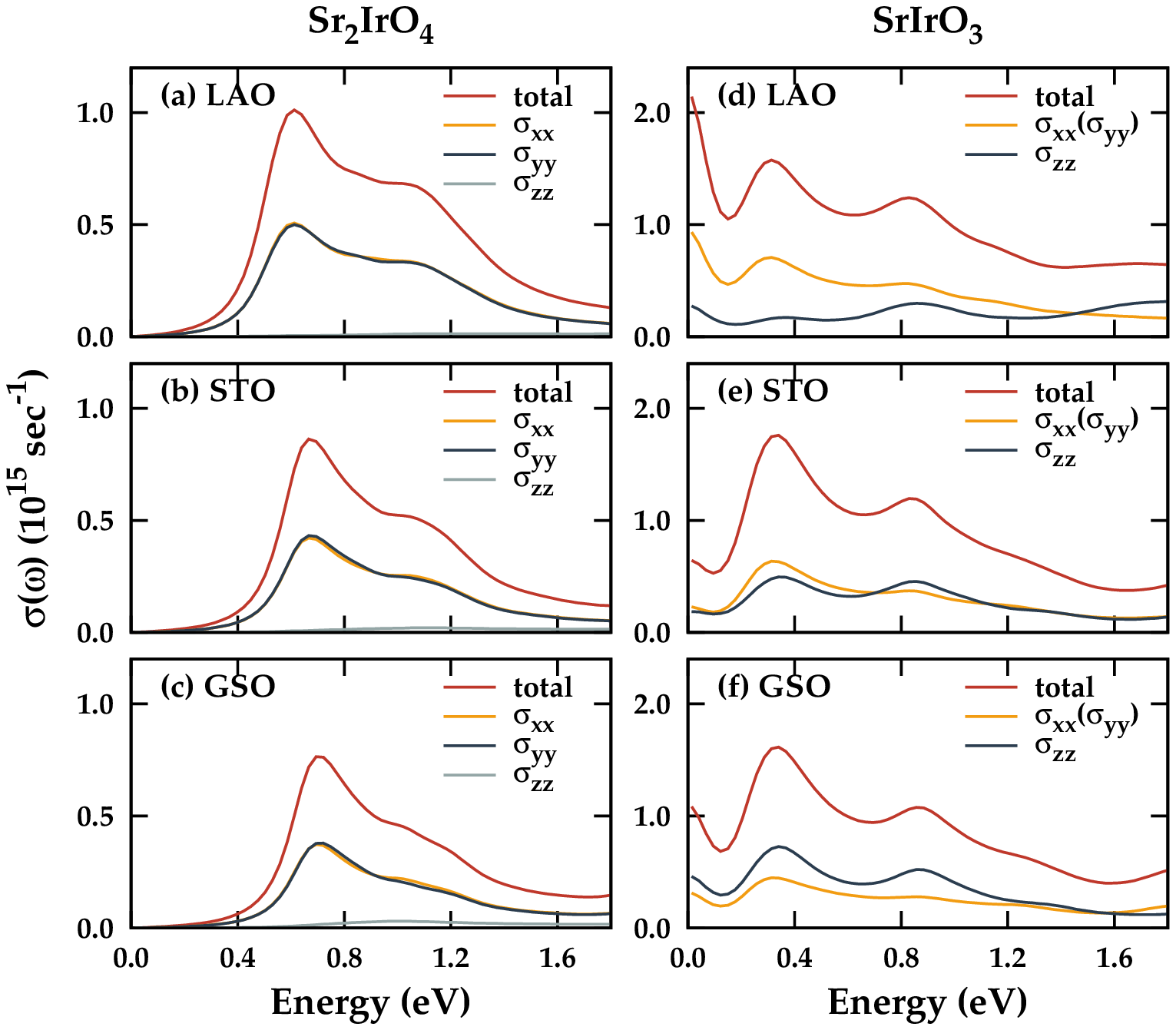}
\caption{
\textbf{Calculated optical conductivities for Sr$_2$IrO$_4$ and SrIrO$_3$ with different substrates.
}
Each case is plotted with diagonal components of conductivity tensor.
}
\label{sfig1}
\end{figure}

 Epitaxially strained 214 systems have tetragonal structure with orthorhombic magnetic symmetry.
The optical conductivity tensor has only nonzero diagonal components ($\sigma_{xx}$, $\sigma_{yy}$,
and $\sigma_{zz}$) with strong two-dimensional character.
Thus there is negligible contribution from $\sigma_{zz}$ component (Fig.~\ref{sfig1}(a)-(c)).

113 systems have tetragonal structure with nonmagnetic ground state.
The overall contributions of the IP and OOP components are systematically changed,
as shown in Fig.~\ref{sfig1}(d)-(f).

\section{Partial density of states}

\begin{figure}[h!]
\includegraphics[angle=270,width=85mm]{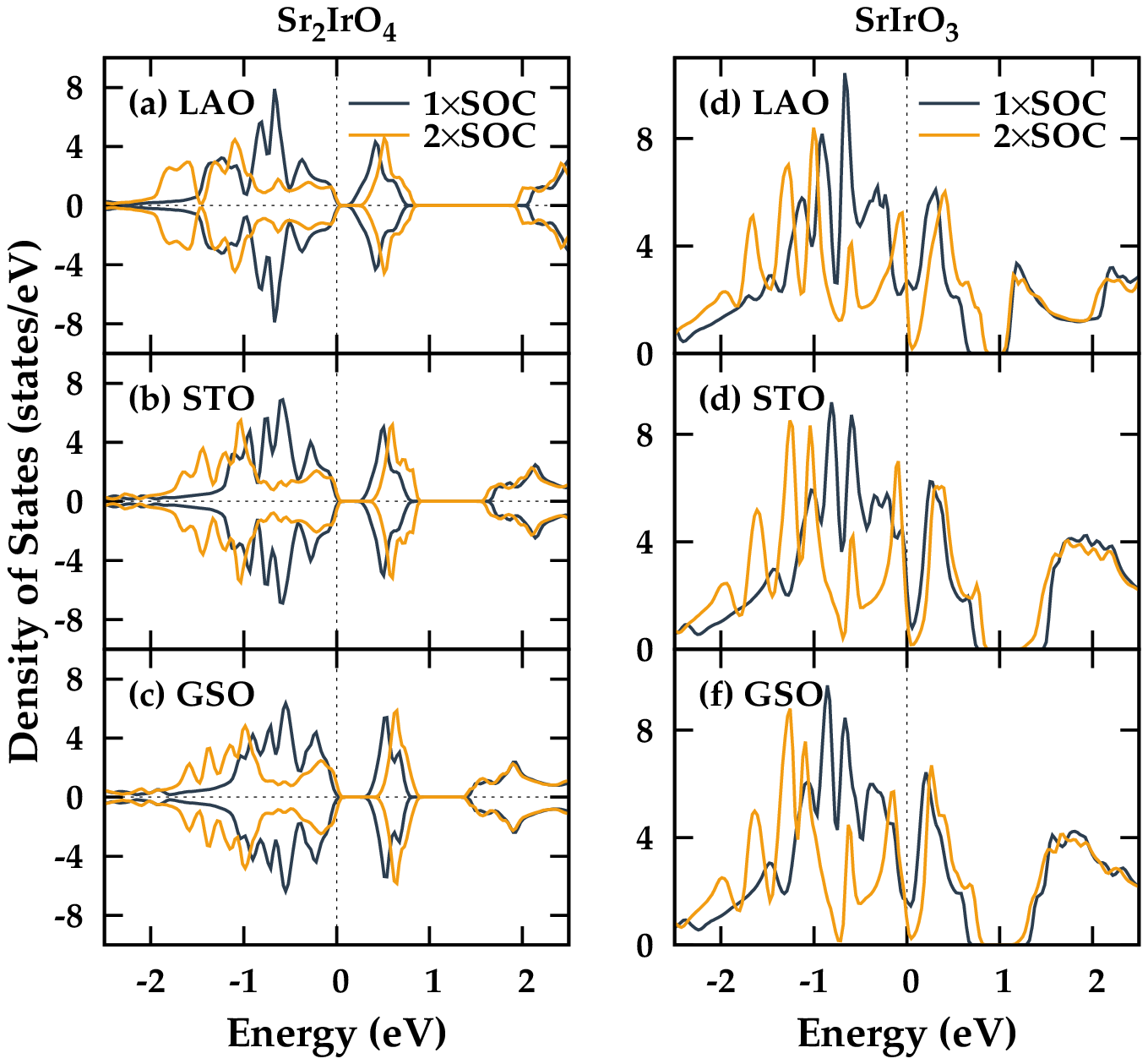}
\caption{
\textbf{Ir-$d$ partial densities of states (DOSs) of Sr$_2$IrO$_4$ and SrIrO$_3$ for different substrates.
}
}
\label{sfig2}
\end{figure}

 The partial densities of states (DOSs) of Ir-$d$ are shown in Fig.~\ref{sfig2}.
In 214 system, one can clearly see the enhancement of localized character upon tensile strain,
as revealed by the sharpening of the DOS.
Also, for the 2$\times$SOC case, clear shift-down of $d$-states is shown,
as expected from the enhanced separation of $J_{eff}$=1/2 and $J_{eff}$=3/2 bands.
Overall behaviors in 113 system are similar to those in 214 system.

\section{Magnetic moment deviation at around $J_{eff}$=1/2 upon strain}

Following the model by Jackeli and Khaliullin,\cite{Jackeli09}
$J_{eff}$=1/2 isospin doublet is expressed as:
\begin{equation}\label{doublet}
  {| \tilde{\pm} \rangle } = \pm \sin \theta {|0,\pm \rangle} \mp
	\cos \theta {| \pm 1, \mp \rangle},
\end{equation}
where $\theta$ denotes parameterized angle incorporating tetragonal crystal field
splitting ($\Delta=E_{xy}-E_{yz/zx}$)
and SOC ($\lambda$) as $\tan (2\theta)=2 \sqrt{2} \lambda / (\lambda - 2 \Delta)$.
The expected spin and orbital magnetic moments are given by
\begin{align}
 &\text{OOP} :& {\mu}_{S} =&  \cos^2\theta-\sin^2\theta, \\
 &\text{OOP} :& {\mu}_{O} =&  \cos^2\theta, \\
 &\text{IP}  :& {\mu}_{S} =&  \sin^2\theta, \\
 &\text{IP}  :& {\mu}_{O} =&  \sqrt{2}\cos\theta \sin\theta ,
\end{align}
where $\mu_S$ and $\mu_O$ in OOP (IP) are calculated with
${\langle2 s_z \rangle}$ (${\langle2 s_x \rangle}$) and
$-{\langle l_z \rangle }$ ($-{\langle l_x \rangle }$), respectively.
We can get following $\mu_{O}/\mu_{S}$ ratios
\begin{align}
 {\mu_{O}/\mu_{S}}  & =  \frac{1}{1-\tan^2\theta}
       =   \frac{4(1-\delta)^{-1}}{\sqrt{\delta^2-2\delta+9}+\delta-1} ~\text{: OOP}\\
 {\mu_{O}/\mu_{S}} & =  \frac{\sqrt{2}} {\tan\theta}
   = \frac{4}{\sqrt{\delta^2-2\delta+9}+\delta-1}  ~\text{: IP},
\end{align}
where $\delta = \frac{2\Delta}{\lambda}$ and $\delta<1$.
Cubic case corresponds to $\delta=0$ and then ${\mu_{O}/\mu_{S}}$ becomes $2$ for any direction.
When the strain is applied, there is a deviation ($\delta$).
Positive (negative) $\delta$ always gives rise to the decrement (increment) of
${\mu_{O}/\mu_{S}}$ from $2$ for IP.
In contrast, ${\mu_{O}/\mu_{S}}$ for OOP increases (decreases) with positive (negative) $\delta$.
Unless other correlation effects modify local electronic structure,
$\delta$ can be more negative when the tensile strain becomes stronger.
${\mu_{O}/\mu_{S}}$ ratio is expected to be larger (smaller) in the IP (OOP) case.
Note that each tensile and compressive strain corresponds to $\delta < 0$ and $\delta > 0$.


\begin{thebibliography}{99}



\bibitem{Rondinelli11}
    J. M. Rondinelli, and N. A. Spaldin.
    Structure and properties of functional oxide thin films: insights from electronic‐structure calculations.
    Adv. Mater. {\bf 23}, 3363 (2011).
\bibitem{Schlom07}
    D. G. Schlom, L.-Q. Chen, C.-B. Eom, K. R. Rabe, S. K. Streiffer, and J.-M. Tricone.
    Strain tuning of ferroelectric thin films.
    Annu. Rev. Mater. Res. {\bf 37}, 589 (2007).
\bibitem{BJKim08}
    B. J. Kim, H. Jin, S. J. Moon, J. Y. Kim, B. G. Park, C. S. Leem, J. Yu, T. W. Noh,
    C. Kim, S. J. Oh, J. H. Park, V. Durairaj, G. Cao, and E. Rotenberg.
    Novel J$_{eff}$=1/2 Mott state induced by relativistic spin-orbit coupling in Sr$_2$IrO$_4$.
    Phys. Rev. Lett. {\bf 101}, 076402 (2008).
\bibitem{Moon08}
    S. J. Moon, H. Jin, K. W. Kim, W. S. Choi, Y. S. Lee, J. Yu, G. Cao, A. Sumi,
    H. Funakubo, C. Bernhard, and T. W. Noh.
    Dimensionality-controlled insulator-metal transition and correlated metallic state
    in 5d transition metal oxides Sr$_{n+1}$Ir$_n$O$_{3n+1}$ (n= 1, 2, and $\infty$).
    Phys. Rev. Lett. {\bf 101}, 226402 (2008).
\bibitem{BJKim09}
    B. J. Kim, H. Ohsumi, T. Komesu, S. Sakai, T. Morita, H. Takagi, and T. Arima.
    Phase-sensitive observation of a spin-orbital Mott state in Sr$_2$IrO$_4$.
    Science {\bf 323}, 1329 (2009).
\bibitem{Jackeli09}
    G. Jackeli and G. Khaliullin.
    Mott insulators in the strong spin-orbit coupling limit: from Heisenberg
    to a quantum compass and Kitaev models.
    Phys. Rev. Lett. {\bf 102}, 017205 (2009).
\bibitem{Sala14}
    M. M. Sala, K. Ohgushi, A. Al-Zein, Y. Hirata, G. Monaco, and M. Krisch.
    CaIrO$_3$: a spin-orbit Mott insulator beyond the j$_{eff}$=1/2 ground state.
    Phys. Rev. Lett. {\bf 112}, 176402 (2014).
\bibitem{Liu13}
    J. Liu, J.-H. Chu, C. Rayan Serrao, D. Yi, J. Koralek, C. Nelson, C. Frontera,
    D. Kriegner, L. Horak, E. Arenholz, J. Orenstein, A. Vishwanath, X. Marti, and R. Ramesh.
    Tuning the electronic properties of J$_{eff}$=1/2 correlated semimetal in epitaxial perovskite SrIrO$_3$.
    arXiv:1305.1732 (2013).
\bibitem{Li15}
   Y. Li, K. Foyevtsova, H. O. Jeschke, and Roser Valent\'{i}.
   Analysis of the optical conductivity for A$_2$IrO$_3$ (A= Na, Li) from first principles.
   Phys. Rev. B {\bf 91}, 161101(R) (2015).
\bibitem{Serrao13}
    C. Rayan Serrao, J. Liu, J. T. Heron, G. Singh-Bhalla, A. Yadav, S. J. Suresha,
    R. J. Paull, D. Yi, J.-H. Chu, M. Trassin, A. Vishwanath, E. Arenholz, C. Frontera,
    J. \v{Z}elezn\'{y}, T. Jungwirth, X. Marti, and R. Ramesh.
    Epitaxy-distorted spin-orbit Mott insulator in Sr$_2$IrO$_4$ thin films.
    Phys. Rev. B {\bf 87}, 085121 (2013).
\bibitem{Nichols13}
    J. Nichols, J. Terzic, E. G. Bittle, O. B. Korneta, L. E. De Long, J. W. Brill,
    G. Cao, and S. S. A. Seo.
    Tuning electronic structure via epitaxial strain in Sr$_2$IrO$_4$ thin films.
    Appl. Phys. Lett. {\bf 102}, 141908 (2013).
\bibitem{Lupascu14}
    A. Lupascu, J. P. Clancy, H. Gretarsson, Z. Nie, J. Nichols, J. Terzic, G. Cao, S. S. A. Seo, Z. Islam, M. H. Upton, J. Kim, D. Casa, T. Gog, A. H. Said, V. M. Katukuri, H. Stoll, L. Hozoi, J. van den Brink, and Y.-J. Kim.
    Tuning magnetic coupling in Sr$_2$IrO$_4$ thin films with epitaxial strain.
    Phys. Rev. Lett. {\bf 112}, 147201 (2014).
\bibitem{Gruenewald14}
   J. H. Gruenewald, J. Nichols, J. Terzic, G. Cao, J. W. Brill, and S. S. A. Seo.
   Compressive strain-induced metal–insulator transition in orthorhombic SrIrO$_3$ thin films.
   J. Mater. Res. {\bf 29}, 2491 (2014).
\bibitem{Moon09}
   S. J. Moon, H. Jin, W. S. Choi, J. S. Lee, S. S. A. Seo, J. Yu, G. Cao, T. W. Noh, and Y. S. Lee.
   Temperature dependence of the electronic structure of the J$_{eff}$=1/2 Mott insulator
   Sr$_2$IrO$_4$ studied by optical spectroscopy
   Phys. Rev. B {\bf 80}, 195110 (2009).
\bibitem{Crawford94}
   M. K. Crawford, M. A. Subramanian, R. L. Harlow, J. A. Fernandez-Baca, Z. R. Wang, and D. C. Johnston.
   Structural and magnetic studies of Sr$_2$IrO$_4$.
   Phys. Rev. B {\bf 49}, 9198 (1994).
\bibitem{Zhao08}
   J. G. Zhao, L. X. Yang, Y. Yu, F. Y. Li, R. C. Yu, Z. Fang, L. C. Chen, and C. Q. Jin.
   High-pressure synthesis of orthorhombic SrIrO$_3$ perovskite and its positive magnetoresistance.
   J. Appl. Phys. {\bf 103}, 103706 (2008).
\bibitem{Zhang13}
   H. Zhang, K. Haule, and D. Vanderbilt.
   Effective J=1/2 insulating state in Ruddlesden-Popper iridates: an LDA+DMFT study.
   Phys. Rev. Lett. {\bf 111}, 246402 (2013).
\bibitem{suppl}
   See Supplement Material for the component-wise optical conductivity data, partial density of states,
   and derivation of Eq. \eqref{mu_IP} and \eqref{mu_OOP}.
\bibitem{Katukuri12}
   V. M. Katukuri, H. Stoll, J. v. d. Brink, L. Hozoi.
   Ab initio determination of excitation energies and magnetic couplings in correlated quasi-two-dimensional iridates.
   Phys. Rev. B {\bf 85}, 220402(R) (2012).
\bibitem{PLiu15}
    P. Liu, S. Khmelevskyi, B. Kim, M. Marsman, D. Li, X.-Q. Chen, D. D. Sarma, G. Kresse, and C. Franchini.
    Anisotropic magnetic couplings and structure-driven canted to collinear transitions in Sr$_2$IrO$_4$
    by magnetically constrained noncollinear DFT.
    Phys. Rev. B \textbf{92,} 054428 (2015).
\bibitem{Boseggia13}
    S. Boseggia, R. Springell, H. C. Walker, H. M. R{\o}nnow, Ch. R\"{u}egg, H. Okabe, M. Isobe,
    R. S. Perry, S. P. Collins, and D. F. McMorrow.
    Robustness of basal-plane antiferromagnetic order and the J$_{eff}$=1/2 state in single-layer iridate
    spin-orbit Mott insulators.
    Phys. Rev. Lett. {\bf 110}, 117207 (2013).
\bibitem{BHKim12}
    B. H. Kim, G. Khaliullin, and B. I. Min.
    Magnetic couplings, optical spectra, and spin-orbit exciton in 5d electron Mott insulator Sr$_2$IrO$_4$.
    Phys. Rev. Lett. {\bf 109}, 167205 (2012).
\bibitem{Sala14_2}
    M. M. Sala, S. Boseggia, D. F. McMorrow, and G. Monaco.
    Resonant X-ray scattering and the j$_{eff}$=1/2 electronic ground state in iridate perovskites.
    Phys. Rev. Lett. {\bf 112}, 026403 (2014).
\bibitem{anisotropy}
    	The different responses upon strain between the IP and OOP cases can also be viewed
	as increased anisotropy in the electronic structure.
	Isotropic $J_{eff}=1/2$ ground state becomes anisotropic due to the crystal field
    	$\delta$ coming from the strain, and the relatively higher change
	in the electronic structure shown in the OOP case can be interpreted
	as stronger dependence on tetragonal distortion $\delta$ of the electronic structure,
    	which has been shown for $\mu_{O}/\mu_{S}$ behaviors
	(See Eq. \eqref{mu_IP} and \eqref{mu_OOP}).
\bibitem{Zeb12}
   M. A. Zeb and H.-Y. Kee.
   Interplay between spin-orbit coupling and Hubbard interaction in SrIrO$_3$ and related Pbnm perovskite oxides.
   Phys. Rev. B {\bf 86}, 085149 (2012).
\bibitem{Kotliar04}
   G. Kotliar and D. Vollhardt.
   Strongly correlated materials: insights from dynamical mean-field theory.
   Phys. Today {\bf 57} (3), 53 (2004).
\bibitem{Biswas14}
    A. Biswas, K.-S. Kim, and Y. H. Jeong.
    Metal insulator transitions in perovskite SrIrO$_3$ thin films.
    J. Appl. Phys. {\bf 116}, 213704 (2014).
\bibitem{Nie15}
    Y. F. Nie, P. D. C. King, C. H. Kim, M. Uchida, H. I. Wei, B. D. Faeth, J. P. Ruf, J. P. C. Ruff, L. Xie, X. Pan, C. J. Fennie, D. G. Schlom, and K. M. Shen. Interplay of spin-orbit interactions, dimensionality, and octahedral rotations in semimetallic SrIrO$_3$.
    Phys. Rev. Lett. {\bf 114}, 016401 (2015).
\bibitem{Liu15}
    Z. T. Liu, M. Y. Li, Q. F. Li, J. S. Liu, D. W. Shen, W. Li, H. F. Yang, Q. Yao, C. C. Fan, X. G. Wan, L. X. You, and Z. Wang.
    Breakdown of the J$_{eff}$=1/2, 3/2 picture in epitaxial perovskite SrIrO$_3$ thin films.
    arXiv:1501.00654.
\bibitem{Zhang15}
   L. Zhang, Q. Liang, Y. Xiong, B. Zhang, L. Gao, H. Li, Y. B. Chen, J. Zhou, S.-T. Zhang, Z.-B. Gu, S.-H. Yao, Z. Wang, Y. Lin, and Y.-F. Chen.
   Tunable semimetallic state in compressive-strained SrIrO$_3$ films revealed by transport behavior.
   Phys. Rev. B {\bf 91} 035110 (2015).
\bibitem{Carter12}
    J.-M. Carter, V. V. Shankar, M. A. Zeb, and H.-Y. Kee.
    Semimetal and topological insulator in perovskite iridates.
    Phys. Rev. B {\bf 85}, 115105 (2012).
\bibitem{HSKim15}
    H.-S. Kim, Y. Chen, and H.-Y. Kee.
    Surface states of perovskite iridates AIrO$_3$: signatures of a topological crystalline metal with nontrivial Z$_2$ index.
    Phys. Rev. B {\bf 91}, 235103 (2015).
\bibitem{JLiu15}
    J. Liu, D. Kriegner, L. Horak, D. Puggioni, C. Rayan Serrao, R. Chen, D. Yi, C. Frontera, V. Holy, A. Vishwanath, J. M. Rondinelli, X. Marti, and R. Ramesh,
    Lifting Dirac nodes in topological semimetallic perovskite SrIrO$_3$ through epitaxial constraint.
    arXiv:1506.03559
\bibitem{Matsuno15}
   J. Matsuno, K. Ihara, S. Yamamura, H. Wadati, K. Ishii, V. V. Shankar, H.-Y. Kee, and H. Takagi.
   Engineering a spin-orbital magnetic insulator by tailoring superlattices.
   Phys. Rev. Lett. {\bf 114}, 247209 (2015).
\bibitem{Weinert82}
    M. Weinert, E. Wimmer, and A. J. Freeman.
    Total-energy all-electron density functional method for bulk solids and surfaces.
    Phys. Rev. B {\bf 26}, 4571(1982).
\bibitem{Jansen84}
	H. J. F. Jansen and A. J. Freeman.
    Total-energy full-potential linearized augmented-plane-wave method for bulk solids: Electronic and structural properties of tungsten.
    Phys. Rev. B {\bf 30}, 561 (1984).
\bibitem{Wien2k}
    P. Blaha, K. Schwarz, G. Madsen, D. Kvasnicka, and J. Luitz, \emph{WIEN2k, An Augmented Plane Wave Plus Local Orbitals Program for Calculating Crystal Properties ISBN 3-9501031-1-1}. Karlheinz Schwarz, Techn. Universit\"{a}t Wien, Austria, 2001.
\bibitem{Moreira02}
   I. de P. R. Moreira, F. Illas, and R. L. Martin.
   Effect of Fock exchange on the electronic structure and magnetic coupling in NiO.
   Phys. Rev. B {\bf 65}, 155102 (2002).
\bibitem{Tran06}
   F. Tran, P. Blaha, K. Schwarz, and P. Nov\'{a}k.
   Hybrid exchange-correlation energy functionals for strongly correlated electrons: applications to transition-metal monoxides.
   Phys. Rev. B {\bf 74}, 155108 (2006).
\bibitem{Ernzerhof99}
   M. Ernzerhof and G. E. Scuseria.
   Assessment of the Perdew–Burke–Ernzerhof exchange-correlation functional.
   J. Chem. Phys. {\bf 110}, 5029 (1999).
\bibitem{Adamo99}
   C. Adamo and V. Barone.
   Toward reliable density functional methods without adjustable parameters: the PBE0 model.
   J. Chem. Phys. {\bf 110}, 6158 (1999).
\bibitem{Franchini14}
   C. Franchini.
   Hybrid functionals applied to perovskites.
   J. Phys. Condens. Matter {\bf 26}, 253202 (2014).
\bibitem{Gangopadhyay15}
   S. Gangopadhyay and W. E. Pickett.
   Spin-orbit coupling, strong correlation, and insulator-metal transitions:
   the J$_{eff}$=3/2 ferromagnetic Dirac-Mott insulator Ba$_2$NaOsO$_6$.
   Phys. Rev. B {\bf 91}, 045133 (2015).
\bibitem{Miao14}
   L. Miao, H. Xu, and Z. Q. Mao.
   Epitaxial strain effect on the J$_{eff}$=1/2 moment orientation in Sr$_2$IrO$_4$ thin films.
   Phys. Rev. B {\bf 89}, 035109 (2014).
\bibitem{MJKim15}
    M. J. Kim and B. I. Min, \emph{unpublished} (2015).
\bibitem{Draxl06}
   C. Ambrosch-Draxl and J. O. Sofo.
   Linear optical properties of solids within the full-potential linearized augmented planewave method.
   Comput. Phys. Commun. {\bf 175}, 1 (2006).
\end{thebibliography}

\begin{thebibliography}{99}
\bibitem{Jackeli09}
    G. Jackeli and G. Khaliullin.
    Mott insulators in the strong spin-orbit coupling limit: from Heisenberg
    to a quantum compass and Kitaev models.
    Phys. Rev. Lett. {\bf 102}, 017205 (2009).
\end{thebibliography}
\end{document}